\documentclass[english]{elsart}
\usepackage{times}
\usepackage[T1]{fontenc}
\usepackage[latin1]{inputenc}
\usepackage{graphicx}

\makeatletter

\providecommand{\LyX}{L\kern-.1667em\lower.25em\hbox{Y}\kern-.125emX\@}
\usepackage{babel}
\makeatother
\begin{document}
\begin{frontmatter}

\title{Theoretical Study of Electronic Structure and Superconductivity in
$Nb_{1-x}B_{2}$ Alloys}

\author{P. Jiji Thomas Joseph and Prabhakar P. Singh}

%\ead{email ppsingh@phy.iitb.ac.in}

\address{Department of Physics. Indian Institute of Technology, Powai, Mumbai-400076,
India }

\begin{abstract}
Using the Korringa-Kohn-Rostoker coherent-potential approximation
in the atomic-sphere approximation (KKR-ASA CPA) we have studied the
changes in the electronic structure and the superconducting transition
temperature $T_{c}$ in $Nb_{1-x}B_{2}$ alloys as a function of $x$.
We find that the variation in the electronic structure of $Nb_{1-x}B_{2}$
alloys as a function of $x$ is consistent with the rigid-band model.
However, the variation of $T_{c}$, obtained using the Allen-Dynes
equation within the Gaspari-Gyorffy formalism to estimate the electron-phonon
matrix elements, does not follow the expected trend. We associate
this disagreement to the use of a constant $\omega _{rms}$ in the
Allen-Dynes equation over the whole range of vacancy concentration,
thereby indicating the importance of lattice dynamical effects in
these systems. 
\end{abstract}
\end{frontmatter}
%%\maketitle

\section{INTRODUCTION}

The discovery of superconductivity in $MgB_{2}$ and the fact that
the carbides and the nitrides of $Nb$ show superconductivity have
renewed interest in finding superconductivity in the borides of $Nb$,
in particular, the diborides. However, the search for superconductivity
in $NbB_{2}$ and $NbB_{2}$-based alloys is not new. Earlier work
by Zeigler $et\, al$ \cite{zeigler} and Hulm $et\, al$ \cite{hulm}
showed $B$ deficient $NbB_{2}$, namely $NbB_{1.94}$ to be superconducting
with superconducting transition temperature, $T_{c}$, of around $1.2\, K$,
while Schriber $et\, al$ \cite{schriber} found $NbB_{x}$ (synthesized
under pressure) with $x\sim 2$ to be superconducting with $T_{c}$
of $9.2\, K$. 

Recently, Yamamoto $et\, al$ \cite{yamamoto} and Kotegawa $et\, al$
\cite{kotegawa} have reported superconductivity in $Nb_{1-x}B_{2}$
alloys when synthesized under pressure for $x$ ranging from $0$
to $0.48$. They find that $Nb_{1-x}B_{2}$ samples become superconducting
for $x=0.04$, and the $T_{c}\textrm{ }$ increases to around $9.2\, K$
for $x=0.24$, thereafter $T_{c}$ remains essentially unchanged up
to $x=0.48$. 

In an effort to theoretically understand the changes in the electronic
structure of $Nb_{1-x}B_{2}$ and $NbB_{1.94}$ alloys and their superconducting
properties as vacancies are introduced either in the $Nb$ plane or
the $B$ plane, we have carried out first-principles electronic structure
calculations of these alloys as a function of vacancy concentration.
We have used Korringa-Kohn-Rostoker coherent-potential approximation
\cite{pps_cpa,faulkner} in the atomic-sphere approximation (KKR-ASA
CPA) method for taking into account the effects of disorder, Gaspari-Gyorffy
formalism \cite{gaspari} for calculating the electron-phonon coupling
constant $\lambda $, and Allen-Dynes equation \cite{allen1} for
calculating $T_{c}$ in $Nb_{1-x}B_{2}$ and $NbB_{1.94}$ alloys. 

Here, we like to point out that the recent theoretical work of Shein
$et$ $al$ \cite{shein} to model disordered $Nb_{0.75}B_{2}$ alloy using a 12-atom
supercell may not be adequate. This can be seen as follows. Each metal
atom in $AlB_{2}$-type diborides has $D_{6h}$ symmetry and thus
has 12 $B$ neighbors at the vertices of a hexagonal prism. In addition,
the central metal atom also sees $8$ others through the faces of
the $B_{12}$ prism at a distance comparable with the metallic diameter.
Thus a minimum of $24$-atom supercell would have been more appropriate.
However, their conclusion that the superconductivity in $Nb_{1-x}B_{2}$
alloys is governed more by the changes that take place in the lattice
properties when vacancies are introduced in the metal plane of the
diborides, is consistent with our results. Before describing our results,
we provide some of the computational details of the present approach.

\section{Computational Details}

\subsection{The densities of states}

The charge self-consistent electronic structure of $Nb_{1-x}B_{2}$
and $NbB_{1.93}$ alloys has been calculated using the KKR-ASA CPA
method with $x$ ranging from $0$ to $0.48.$ The lattice constants
for $Nb_{1-x}B_{2}$ alloys are taken from the experimental work of
Yamamoto $et\, al$\cite{yamamoto}. To determine the electronic structure
of $NbB_{2}$ using the present approach, we have used the lattice
constants $a$ and $c$ from our earlier full-potential linear muffin-tin
orbital calculations\cite{pps_nbb2}. The absence of $Nb$ and $B$ atoms
in the $Nb$ and $B$ planes, respectively, were modelled by introducing
empty spheres at those sites. The disorder was accounted for by the
coherent-potential approximation \cite{soven,taylor,kirk}, which
has been very successful in describing reliably many physical properties
of disordered alloys \cite{faulkner}. The exchange-correlation potential
was parametrized as suggested by Perdew-Wang within the generalized
gradient approximation \cite{perdew1,perdew2}. The Brillouin zone
(BZ) integration was carried out with $1215$ \textbf{k}-points in
the irreducible part of the BZ. For densities of states (DOS) calculations,
we added a small imaginary component of $1$ $mRy$ to the energy
and used $4900$ \textbf{k}-points in the irreducible part of the
BZ. For valence states the angular momentum cut-off was kept as $l_{max}=3$,
and the core states were recalculated. For $Nb$ and $B$ the ratio of the 
atomic sphere radius 
to the Wigner-Seitz radius  was kept fixed at 1.294 and 0.747, respectively.
  The empty spheres inherited the radii of the 
corresponding atomic spheres. The present calculations were
carried out using the scalar-relativistic Schroedinger equation. The
Green's function is calculated in a complex plane with a $20$ point
Gaussian quadrature.

\subsection{The electron-phonon coupling and the superconducting transition temperature}

The superconducting transition temperature $T_{c}$ was calculated
with the Allen-Dynes equation,

\begin{equation}
T_{c}=\frac{\omega _{ln}}{1.2}\exp \left[-\frac{1.04(1+\lambda )}{\lambda -\mu ^{*}(1+0.62\lambda )}\right].\label{tc}\end{equation}
 In our calculation, the value of the phonon frequency $\omega _{ln}$
for $NbB_{2}$ was taken from Ref. \cite{pps_nbb2}, $\mu ^{*}$ was set equal
to 0.09, and the electron-phonon coupling constant $\lambda $ was
calculated using the Gaspari-Gyorffy formalism with the charge self-consistent
potentials of $Nb_{1-x}B_{2}$ and $NbB_{1.93}$ alloys obtained with
the KKR-ASA CPA method. This entails writing $\lambda $ in terms
of Hopfield parameter, $\eta $, as

\begin{equation}
\lambda =\sum _{i}\frac{\eta _{i}}{M_{i}<\omega _{i}^{2}>},\label{lambda}\end{equation}
 where the sum is over the basis atoms in the primitive cell, $M$
is the atomic mass and $<\omega ^{2}>^{1/2}$ is an average phonon frequency.
The average phonon frequency $<\omega ^{2}>^{1/2}$ is taken to be the root-mean-
square frequency $\omega_{rms} = <\omega^2>^{1/2}$.  
In the Gaspari-Gyorffy formalism the spherically averaged Hopfield
parameter is given by

\begin{equation}
\eta =2\sum _{l}\frac{(l+1)}{(2l+1)(2l+3)}M_{l.l+1}^{2}\frac{N_{l}(E_{F})N_{l+1}(E_{F})}{N(E_{F})}.\label{eta}\end{equation}
 The total DOS at the Fermi energy $E_{F}$, $N(E_{F})$, and the
$l-$resolved DOS, $N_{l}(E_{F})$ as well as the electron-phonon
matrix elements,

\begin{equation}
M_{l,l+1}=-\phi _{l}(E_{F})\phi _{l+1}(E_{F})\left[(D_{l}(E_{F})-l)(D_{l+1}(E_{F})+l+2)+(E_{F}-V(S))S^{2}\right]\label{mllp}\end{equation}
are calculated from the charge self-consistent potentials obtained
with the KKR-ASA CPA method. In Eq. (\ref{mllp}), $\phi _{l}(E_{F})$
is the amplitude of the $l-$th partial wave at the sphere boundary,
$S,$ evaluated at $E_{F},$ $D_{l}(E_{F})$ is the corresponding
logarithmic derivative, and $V(S)$ is the one-electron potential
at $S$.

\section{Results and Discussion}

In the following we describe the results of our calculations for $NbB_{2}$
and its alloys in terms of (i) the densities of states, (ii) the electron-phonon
coupling and (iii) the superconducting transition temperature. We
first describe how the densities of states, the electron-phonon coupling
and the superconducting transition temperature change for low concentrations
of vacancies in either the $Nb$ or the $B$ plane. Then, we describe
the changes in these properties as a function of increasing vacancy
concentration.

\subsection{The densities of states of \textmd{$NbB_{2}$, $Nb_{0.99}B_{2}$}
and \textmd{$NbB_{1.93}$}}

\subsubsection{$NbB_{2}$}

In Fig. 1 we show the calculated total DOS and the partial DOS together
with the $l$-resolved partial DOS of $NbB_{2}$. The overall structure
of the DOS is consistent with the previous work\cite{shein,vajeeston}.
The calculations based on the tight-binding LMTO method yields a value
of total DOS at the Fermi energy equal to $4.65$ states/Ry-cell \cite{vajeeston},
while the corresponding value from the full-potential LMTO calculations
is found to be $4.28$ states/Ry-cell. The present calculations, which
is based on the KKR-ASA method, gives $N(E_{F})$ to be equal to $4.41$
states/Ry-cell. At $E_{F}$ the contributions of $Nb$ and $B$ to
the total DOS is found to be $3.44$ and 0.97 states/Ry-cell, respectively. 

\textbf{}%
\begin{figure}[htbp]
\begin{center}\textbf{\includegraphics[  width=7.4cm,
  height=7.4cm,
  angle=270,
  origin=c]{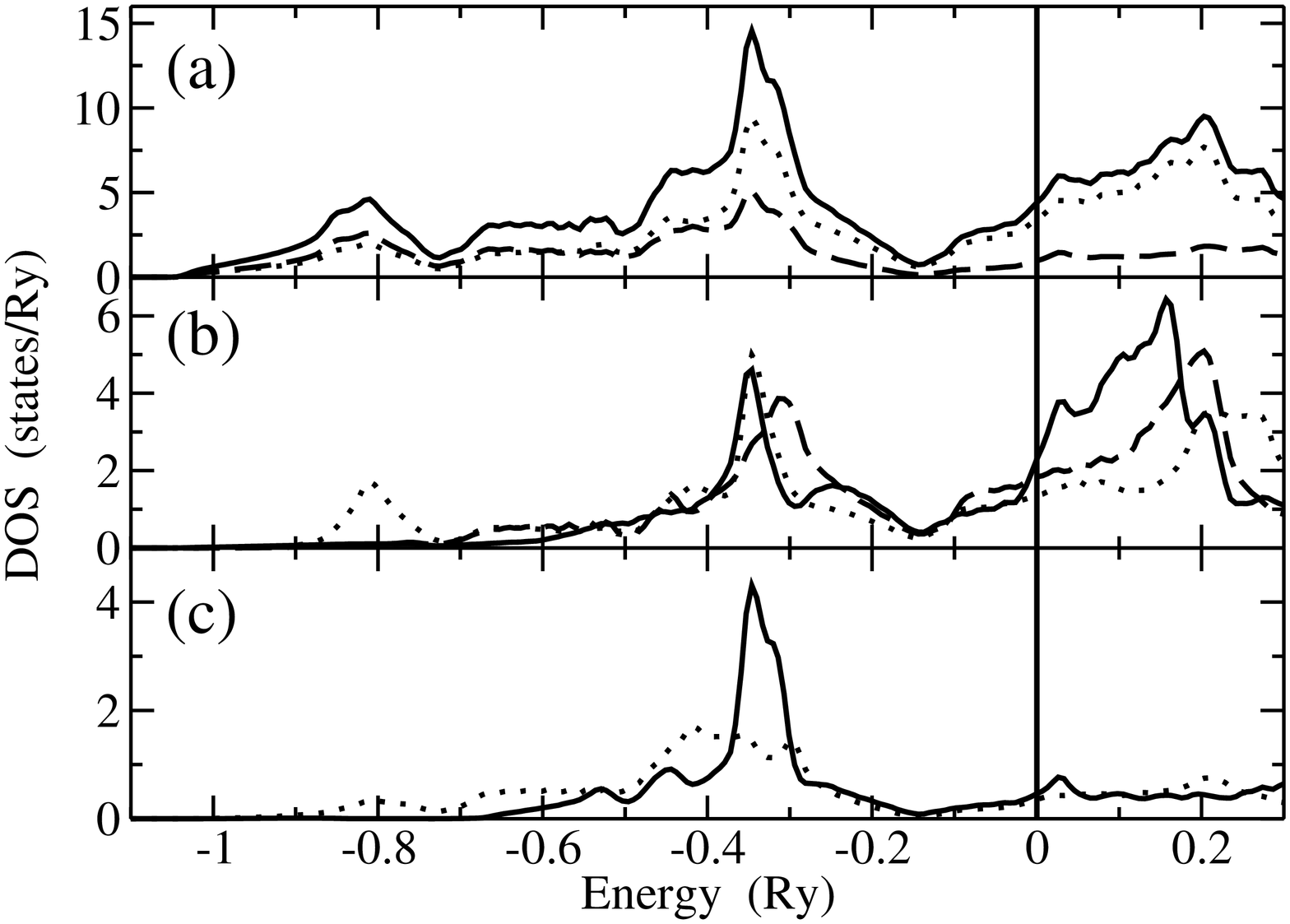}}\end{center}

Fig. 1. (a) The total (solid line) and the partial densities of states
of $Nb$ (dotted line) and $B$ in $NbB_{2}$, calculated with the
KKR-ASA CPA method as described in the text. (b) The various symmetry-resolved
$d$-partial DOS of $Nb$ in $NbB_{2}$ ($d_{xy}$ : dashed line,
$d_{yz}$ : dotted line and $d_{3z^{2}-1}$: full line). Note that
$d_{x^{2}-y^{2}}$ and $d_{xz}$ are degenerate with $d_{xy}$ and
$d_{yz}$, respectively. (c) The symmetry-resolved $p$-partial DOS
of $B$ in $NbB_{2}$ ($p_{z}$ : full line and $p_{x}$ : dotted
line). In this case,  $p_{y}$ is degenerate with $p_{x}$.
\end{figure}

The characteristic features in the DOS of $NbB_{2}$, as shown in
Fig. 1, are $(i)$ a low lying broad peak around $0.90-0.75$ Ry below
$E_{F}$ due to $B$ $s-$ and $Nb\, d-$states, $(ii)$ another broad
peak at around $0.45-0.25$ Ry comprising of $B\, p-$, $Nb\, p-$
and $Nb\, d-$ states, $(iii)$ a relatively constant DOS from $0.7-0.5$
Ry primarily of $B\, p-$ and $Nb\, d-$ states, and $(iv)$ a pseudo-gap
at $0.14$ Ry, which appears in many transition-metal diborides \cite{vajeeston}.
Thus the dominant presence of $Nb\, d-$ states at $E_{F}$ in $NbB_{2}$
makes it quite different from $MgB_{2}$ which has largely $B$ $p-$states
at $E_{F}$.

\subsubsection{\textmd{$Nb_{0.99}B_{2}$ }}

To understand how low concentrations of vacancies in the $Nb$ plane
of $NbB_{2}$ would affect its electronic properties, we have used
KKR-ASA CPA method to calculate the electronic structure of dilute
$Nb_{1-x}B_{2}$ with $x$$=$$0.99$ alloy. In this calculation the
lattice constants were kept the same as that of $NbB_{2}$. In Fig.
2, we show the calculated $l$-resolved partial DOS at the empty sphere
site in $Nb_{0.99}B_{2}$. It is seen that upon $1$\% creation of
vacancy in the $Nb$ plane, the total DOS at $E_{F}$ for $Nb_{0.99}B_{2}$
is reduced by $\sim $$4$\% relative to that of $NbB_{2}$. We find
that the creation of vacancies in the $Nb$ plane do not lead to appreciable
change in the $B$ partial DOS at $E_{F}$. Thus the metal deficiency
introduced in $NbB_{2}$ enhances the ratio of $B$ states to that
of the total DOS at $E_{F}$. 

\textbf{}%
\begin{figure}[htbp]
\begin{center}\textbf{\includegraphics[  width=7.4cm,
  height=7.4cm,
  angle=270,
  origin=c]{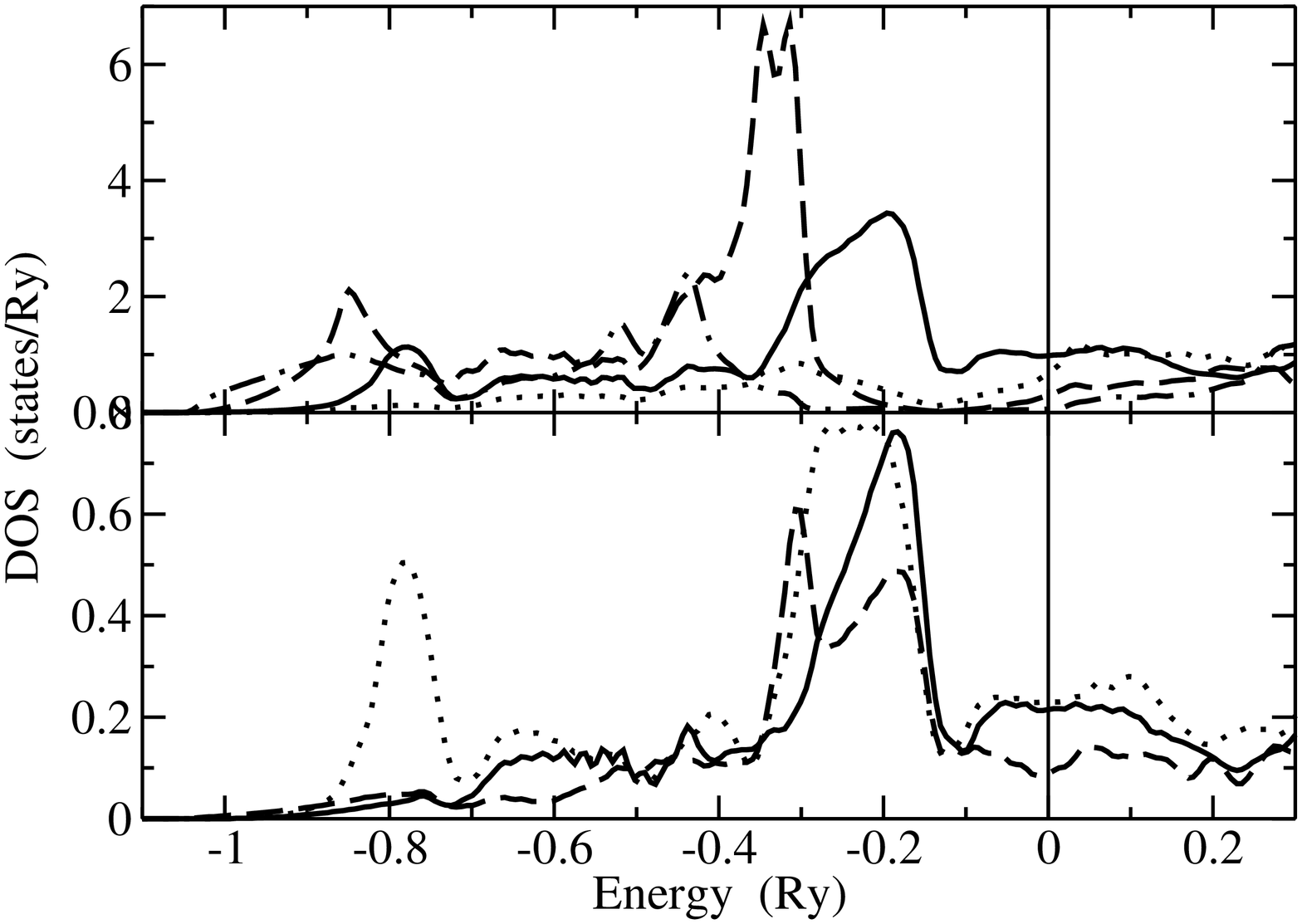}}\end{center}

Fig. 2. Top panel: The $l$-resolved partial densities of states ($s$
: dot-dashed line, $p$ : dashed line, $d$ : solid line, and $f$
: dotted line) of the empty sphere in $Nb_{0.99}B_{2}$ alloy, calculated
with the KKR-ASA CPA method as described in the text. Lower panel:
The symmetry-resolved $d$-partial DOS of the empty sphere in $Nb_{0.99}B_{2}$
($d_{xy}$ : full line, $d_{yz}$ : dotted line and $d_{3z^{2}-1}$:
dashed line). Note that $d_{x^{2}-y^{2}}$ and $d_{xz}$ are degenerate
with $d_{xy}$ and $d_{yz}$, respectively.
\end{figure}

\subsubsection{\textmd{$NbB_{1.93}$ }}

Motivated by the work of Otani $et\, al$ on $NbB_{1.93}$ alloy,
which showed that a depletion of $B$ in $NbB_{2}$ led to an increase
in the lattice parameter $c$ from $6.16$ to $6.25$ a.u. and a decrease
in the lattice parameter $a$ from $5.89$ to $5.84$ a.u., we have
calculated the electronic structure of $NbB_{1.93}$ alloy. The calculation
is carried out by disordering the $B$ sub-lattice with empty spheres
and keeping the $Nb$ sub-lattice ordered. The calculated densities
of states at $E_{F}$ are summarized in Table I, where we have also
listed the lattice constants used in the present work. 

Our analysis of the DOS of $NbB_{1.93}$ shows that the creation of
vacancies in the $B$ plane largely reduces the $d_{3z^{2}-1}$ component
($\sim $12.5\%) of the $Nb$ $d$-states and the $B$ $s$-states
($\sim $8.9\%) at $E_{F}$. The creation of vacancies in the $B$
sub lattice of $NbB_{2}$ thus leads to a major reduction in those
states which are directed along the $c-$axis of the unit cell. 

\textbf{}%
\begin{table}[htbp]
Table I. The lattice constants $a$ and $c$, in atomic units, used
for $NbB_{2}$ and $NbB_{1.93}$ alloys. The calculated partial ($Nb$
and $B$) and total densities of states at $E_{F}$, in states/Ry-cell,
are also given. 

\begin{center}\begin{tabular}{|c|c|c|c|c|c|}
\hline 
&
a&
c&
$N^{Nb}(E_{F})$&
$N^{B}(E_{F})$&
$N^{tot}(E_{F})$\\
\hline
$NbB_{2}$&
5.847&
6.259&
0.497&
3.497&
4.491\\
\hline 
$NbB_{1.93}$&
5.887&
6.161&
0.473&
3.294&
4.241\\
\hline
\end{tabular}\end{center}
\end{table}

\subsection{The electron-phonon coupling and the superconducting transition temperature
of \textmd{$NbB_{2}$, $Nb_{0.99}B_{2}$} and \textmd{$NbB_{1.93}$}
alloys}

The superconducting transition temperatures of $NbB_{2}$, $Nb_{0.99}B_{2}$
and $NbB_{1.93}$ were calculated with the Allen-Dynes equation. The
factor representing the Coulomb interaction $\mu ^{*}$ was 0.09.
For $\omega _{rms}$ and $\omega _{ln}$, we used the values $464\, cm^{-1}$
and $494\, K$, respectively. Our results for the Hopfield parameter
$\eta $, electron-phonon coupling constant $\lambda $ and the superconducting
transition temperature are summarized in Table II. From Table II,
we see that $\eta _{B}$, the Hopfield parameter due to $B$, is relatively
small in comparison to $\eta _{Nb}$, which reflects the fact that
$B$ contribution to the total density of states at $E_{F}$ is small.
Since we use the same $\omega _{rms}$ for $NbB_{2}$, $Nb_{0.99}B_{2}$
and $NbB_{1.93}$ the changes in $\lambda $ are similar to that of
$\eta $. The calculated $T_{c}$ for $NbB_{2}$, $Nb_{0.99}B_{2}$
and $NbB_{1.93}$ are found to be $0.31$K, $0.13$K and $0.004$K,
respectively. In the case of $NbB_{1.93}$ alloy, we find that the
deficiency of $B$ largely affects $\eta _{Nb}$ rather than $\eta _{B}$,
which reduces $T_{c}$ significantly.

\begin{table}[htbp]
Table II. The calculated Hopfield parameters $\eta $, in $mRy/a_{B}^{2}$,
the electron-phonon coupling constant $\lambda $ and the superconducting
transition temperature $T_{c} (K)$ for $NbB_{2}$ and $NbB_{1.93}$ alloys.
Note that 
$\lambda _{tot}$=$\lambda _{Nb}$+ 2$\lambda _{B}$ for $NbB_2$ and
$\lambda _{tot}$=$\lambda _{Nb}$+ 1.93$\lambda _{B}$ for $NbB_{1.93}$.

\begin{center}\begin{tabular}{|c|c|c|c|c|c|c|}
\hline 
&
$\eta _{Nb}$&
$\eta _{B}$&
$\lambda _{Nb}$&
$\lambda _{B}$&
$\lambda _{tot}$&
$T_{c}$\\
\hline 
$NbB_{2}$&
81.7&
18.1&
0.053&
0.121&
0.294&
0.313\\
\hline 
$NbB_{1.93}$&
61.2&
13.9&
0.039&
0.089&
0.212&
0.004\\
\hline
\end{tabular}\end{center}
\end{table}

\subsection{The densities of states of $Nb_{1-x}B_{2}$ alloys }

In Fig. 3, we show the DOS of $Nb_{0.96}B_{2}$, $Nb_{0.84}B_{2}$,
$Nb_{0.76}B_{2}$ and $Nb_{0.52}B_{2}$ alloys using the KKR-ASA CPA
method. The lattice constants for these alloys were taken from Ref.
\cite{yamamoto}. The changes in the DOS that arise in $Nb_{1-x}B_{2}$
alloys as a function of vacancies are $(i)$ the inward movement of
$E_{F}$ with increasing vacancy concentration, leading to a decrease
in the total DOS at $E_{F}$ for $x<0.16$ and an increase for $x>0.16$,
and $(ii)$ the broadening of peaks with increasing vacancy concentration
due to disorder. We find that the movement of $E_{F}$ follows the
trend as suggested by the rigid-band model, and the DOS of $Nb_{0.84}B_{2}$
alloy is similar to $ZrB_{2}$. 

\begin{figure}[htbp]
\begin{center}\includegraphics[  width=7.4cm,
  height=7.4cm,
  angle=270,
  origin=c]{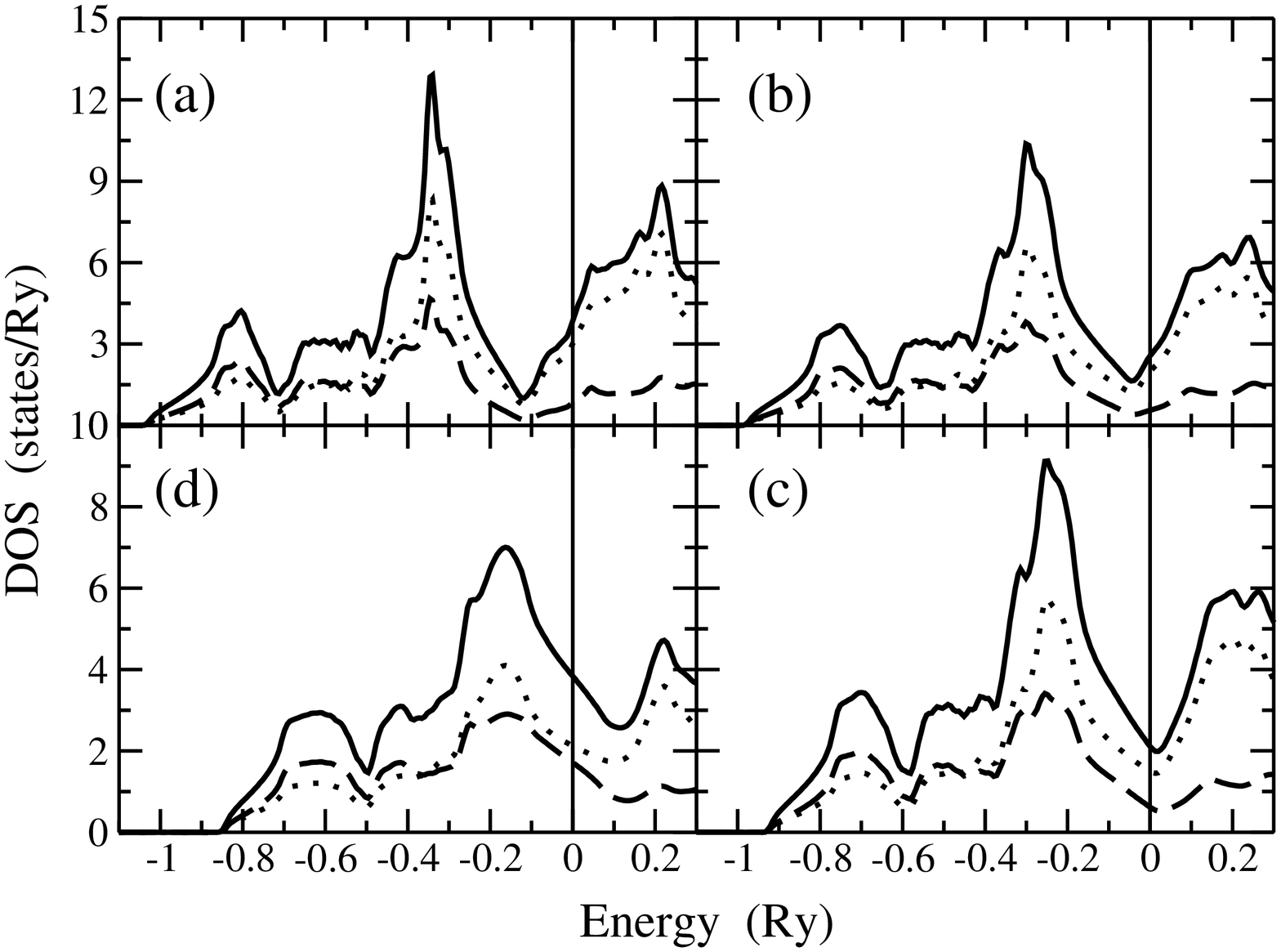}\end{center}

Fig. 3. The total (solid line) and the partial densities of states
of $Nb$ (dotted line) and $B$ (dashed line) in (a) $Nb_{0.96}B_{2}$,
(b) $Nb_{0.84}B_{2}$, (c) $Nb_{0.76}B_{2}$ and (d) $Nb_{0.52}B_{2}$
alloys, calculated with the KKR-ASA CPA method as described in the
text. The vertical line represents the Fermi energy. 
\end{figure}

\subsection{The electron-phonon coupling and the superconducting transition temperature
of $Nb_{1-x}B_{2}$ alloys}

From the resulting electronic structure we have calculated the variation
in the Hopfield parameter, electron-phonon coupling constant and the
superconducting transition temperature of $Nb_{1-x}B_{2}$ alloys
as a function of vacancy concentration. Our results for $\eta $ and
$\lambda $ are summarized in Table III. We find that both $\eta $
and $\lambda $ decrease with increasing vacancy concentration up
to $x\sim 0.25$, thereafter they start to increase. In Fig. 4, we
show the calculated $T_{c}$ as a function of $x$ in $Nb_{1-x}B_{2}$
alloys, together with the experimentally reported values. The parameters
used in the calculation of $T_{c}$ were $\mu ^{*}=0.09$, $\omega _{rms}=400\, cm^{-1}$,
and $\omega _{ln}=494\, K$. The calculated trend in the variation
of $T_{c}$ is in disagreement with the experimental results as shown
in Fig. 4. In our opinion, the disagreement is very likely due to
the use of a constant $\omega _{rms}$ value in the Allen-Dynes equation
for the whole range of $x$ from $0$ to $\textrm{0.48}$ for $Nb_{1-x}B_{2}$
alloys.

In the present
approach, the calculation of $T_{c}$ incorporates two important  approximations,
(i) use of Gaspari-Gyorffy formalism for estimating the electron-phonon
matrix elements, which is known to underestimate the strength of the
coupling, and 
(ii) use of a constant phonon frequency over the whole range of $x$.
Using Gaspari-Gyorffy formalism with concentration-weighted averages of the 
phonon frequencies, we have been able to describe reliably the 
trend in the variation of $T_c$  in many $MgB_2$-based alloys 
\cite{pps_mgtab2,pps_mgalb2,pps_3dmgb2}. 
Since the electron-phonon coupling in $NbB_2$ is dominated by $Nb$ and not by 
$B$, as is the case in $MgB_2$, the Gaspari-Gyorffy formalism may be 
more suitable for $NbB_2$ than for 
$MgB_2$. This can be seen by comparing the $T_c$ as calculated in the 
present approach with that of the linear response calculation of Ref. 
\cite{pps_nbb2}. The lack of agreement between the present 
calculations and the observed $T_c$ in $Nb_{1-x}B_2$ alloys as a function of 
$x$ indicates the importance of 
incorporating the lattice-dynamical effects in  these calculations. 

\begin{table}[htbp]
Table III. The calculated variation of Hopfield parameters $\eta $,
in $mRy/a_{B}^{2}$ and the electron-phonon coupling constant $\lambda $
as a function of $x$ in $Nb_{1-x}B_{2}$ alloys. Note that $\lambda _{tot}$=(1-x)$\lambda _{Nb}$+2$\lambda _{B}$ for $Nb_{1-x}B_{2}$ alloys. 

\begin{center}\begin{tabular}{|c|c|c|c|c|c|}
\hline 
&
$\eta _{Nb}$&
$\eta _{B}$&
$\lambda _{Nb}$&
$\lambda _{B}$&
$\lambda _{tot}$\\
\hline
$NbB_{2}$&
80.3&
21.1&
0.070&
0.158&
0.387\\
\hline 
$Nb_{0.96}B_{2}$&
73.1&
18.2&
0.063&
0.136&
0.334\\
\hline 
$Nb_{0.92}B_{2}$&
65.2&
15.4&
0.057&
0.115&
0.283\\
\hline 
$Nb_{0.84}B_{2}$&
55.9&
12.4&
 0.048&
0.092&
0.227\\
\hline 
$Nb_{0.76}B_{2}$&
47.6&
12.9&
0.041&
0.097&
 0.228\\
\hline 
$Nb_{0.68}B_{2}$&
62.9&
21.0&
0.054&
0.157&
 0.356\\
\hline 
$Nb_{0.60}B_{2}$&
73.3&
27.3&
0.064&
0.205&
0.454\\
\hline 
$Nb_{0.52}B_{2}$&
82.2&
32.1&
0.072&
0.241&
0.526\\
\hline
\end{tabular}\end{center}
\end{table}

\begin{figure}[htbp]
\begin{center}\includegraphics[  width=7.4cm,
  height=7.4cm,
  angle=270,
  origin=c]{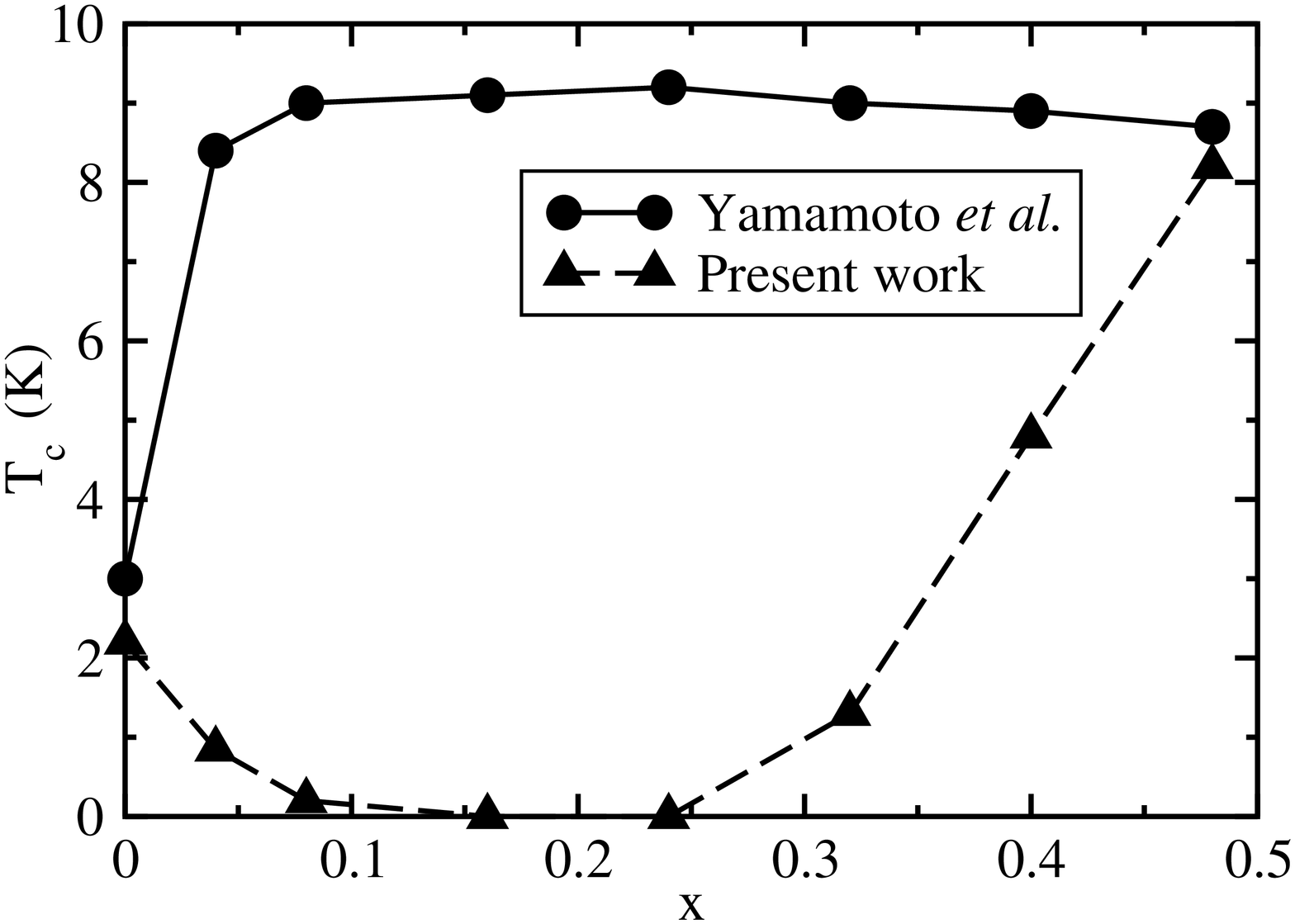}\end{center}

Fig. 4. The calculated (filled triangle) and the observed (filled
circle) superconducting transition temperature $T_c$ of $Nb_{1-x}B_{2}$
alloys as a function of $x$. 
\end{figure}

\section{Conclusions}

We have studied the electronic structure of $NbB_{2}$, $NbB_{1.93}$
and $Nb_{1-x}B_{2}$ alloys using the KKR-ASA CPA method. We find
that the vacancies on the metal plane reduce the number of $Nb$ $d$-states
at $E_{F}$. The changes in the density of states in $Nb_{1-x}B_{2}$
alloys as a function of $x$ reflect a rigid-band picture. The variation
in $T_{c}$ in $Nb_{1-x}B_{2}$ alloys as a function of $x$, calculated
using the Gaspari-Gyorffy formalism disagrees with the observed trend.
In our opinion, this disagreement arises due to the use of a constant
$\omega _{rms}$ in the Allen-Dynes equation over the whole concentration
range in our calculations.

\end{document}